\documentclass[prd,twocolumn,showpacs,preprintnumbers,superscriptaddress,floatfix,nofootinbib]{revtex4-1}
\usepackage{graphicx,,booktabs}
\usepackage{amssymb}
\usepackage{amsmath,txfonts}
\usepackage{graphicx}
\usepackage{dcolumn}
\usepackage{color}
\usepackage{bm}
\usepackage[colorlinks,
citecolor=blue,
anchorcolor=green,
menucolor=orange,
linkcolor=red,
filecolor=red,
runcolor=pink,
urlcolor=blue,
frenchlinks=red]{hyperref}
\usepackage{multirow}

\begin{document}
	
	\title{\boldmath Exploring kaon induced reactions for unraveling the nature of the scalar meson $a_0 (1817)$}

	\author{Xiao-Yun Wang}
	\email{xywang@lut.edu.cn}
	\affiliation{Department of physics, Lanzhou University of Technology,
		Lanzhou 730050, China}
	\affiliation{Lanzhou Center for Theoretical Physics, Key Laboratory of Theoretical Physics of Gansu Province, Lanzhou University, Lanzhou, Gansu 730000, China}
	
  \author{Hui-Fang Zhou}
	\affiliation{Department of physics, Lanzhou University of Technology,
		Lanzhou 730050, China}

	\author{Xiang Liu}
\email{xiangliu@lzu.edu.cn}
\affiliation{School of Physical Science and Technology, Lanzhou University, Lanzhou 730000, China}
\affiliation{Lanzhou Center for Theoretical Physics, Key Laboratory of Theoretical Physics of Gansu Province, Lanzhou University, Lanzhou, Gansu 730000, China}
\affiliation{Key Laboratory of Quantum Theory and Applications of MoE, Lanzhou University,
Lanzhou 730000, China}
\affiliation{Frontiers Science Center for Rare Isotopes, Lanzhou University, Lanzhou 730000, China}
\affiliation{Research Center for Hadron and CSR Physics, Lanzhou University and Institute of Modern Physics of CAS, Lanzhou 730000, China}

	\begin{abstract}
		In this study, we comprehensively investigate the production of isovector scalar meson $a_{0}(1817)$ using the effective Lagrangian approach. Specifically, we employ the Reggeized $t$ channel Born term to calculate the total and differential cross sections for the reaction $K^{-}p \rightarrow a_{0}(1817)\Lambda$. Our analysis reveals that the optimal energy range for detecting the $a_{0}(1817)$ meson lies between $W=3.4$ GeV and $W=3.6$ GeV, where the predicted total cross section reaches a minimum value of 112 nb. Notably, the $t$ channel, as predicted by the Regge model, significantly enhances the differential cross sections, particularly at extreme forward angles. Furthermore, we investigate the Dalitz processes of $2\rightarrow 3$ and discuss the feasibility of detecting the $a_{0}(1817)$ meson in experiments such as J-PARC.
	\end{abstract}

	\maketitle
	
	\section{Introduction}
	
The $a_{0}(1817)$ meson has attracted significant attention in the field of light hadron physics, as its study provides valuable insights into  the intricacies of constructing the light flavor hadron spectroscopy. Recent experimental findings have added to the intrigue surrounding this meson. The BaBar Collaboration, through the $\eta_{c} \rightarrow \eta \pi^{+}\pi^{-}$ reaction, discovered a new state named $a_{0}(1700)$, which has a measured mass of $1704\pm 5 (\text{stat.})\pm 2 (\text{syst.})$ MeV and a width of $\Gamma=110\pm 15 (\text{stat.})\pm 11 (\text{syst.})$ MeV \cite{Babar2021}. Additionally, the BESIII Collaboration observed a state denoted as the $a_{0}(1710)^{0}$ in the $D_{S}^{+} \rightarrow K_{S}^{0}K_{S}^{0}\pi^{+}$ reaction. However, in this detection process, it was not possible to differentiate between the $a_{0}(1710)^{0}$ and $f_{0}(1710)$, leading to a generalization of both states as $S_{0}(1710)$ \cite{BESIII:2021anf}. It was later resolved in a subsequent article using the isospin theorem, which distinguished the isospin $I=1$ state $a_{0}(1710)$ from the isospin $I=0$ state $f_{0}(1710)$ \cite{Dai:2021owu}. Subsequently, the BESIII Collaboration conducted another experiment to study the $a_{0}(1710)^{+}$ state with the quantum numbers $I(J^{P})=1^{+}(0^{+})$. The observation of $a_{0}(1710)^{+}\rightarrow K^{0}_{S}K^{+}$ was made through an investigation of the $D^{+}_{S}\rightarrow K^{0}_{S}K^{+}\pi^{0}$ decay \cite{BESIII:2022npc}. This experiment reported the mass and decay width of the newly discovered meson as $M=1.817\pm0.008 (\text{stat.})\pm 0.020 (\text{syst.})$ GeV and $\Gamma=0.097\pm 0.022 (\text{stat.})\pm 0.015 (\text{syst.})$ GeV, respectively. In accordance with the designation proposed by the Lanzhou group {\it et al.} \cite{Guo:2022xqu}, we adopt the name $a_{0}(1817)$ for this newly discovered isovector scalar meson in our work.

However, there exist discrepancies in the measured mass and decay width of the $a_{0}(1817)$ meson as observed by the BaBar experiment \cite{Babar2021} and the BESIII experiment \cite{Dai:2021owu}. Moreover, due to the limited number of relevant experiments and available experimental data, further observations of the $a_{0}(1817)$ meson in alternative experiments are necessary. These observations would facilitate the measurement of pertinent resonance parameters and provide a more comprehensive understanding of the properties associated with the $a_{0}(1817)$ meson.

Recent research by the Lanzhou group \cite{Guo:2022xqu} has established the significance of the $a_{0}(1817)$ meson as a reference point in the construction of scalar meson families. Its primary decay channels include $\pi\eta(1295)$, $\pi\eta'$, $\pi\eta$, $\pi\eta(1475)$, $\pi b_{1}(1235)$, $K\bar{K}$, and others, with specific details provided in Table \ref{tab:table1}. Theoretical conjectures propose that the $a_{0}(1450)$ and $a_{0}(1817)$ represent the first and second radial excitations, respectively, of the $a_{0}(980)$ meson \cite{Wang:2022pin}. Additionally, it is predicted that the $a_{0}(2115)$ serves as the third radial excitation, contributing significantly to the expanding landscape of the light hadron spectrum \cite{Guo:2022xqu}. Understanding the composition and meson structure of the $a_{0}(1817)$ is crucial, as it sheds light on the structural characteristics of scalar mesons in the realm of light quarks, while also addressing other pertinent issues currently under debate in the field of hadron physics \cite{Oset:2023hyt,Zhu:2022wzk,Zhu:2022guw}. Previous studies have considered the possibility of the $a_{0}(980)$ as a tetraquark candidate \cite{Humanic:2022hpq}, and the $a_{0}(1450)$ as a hybrid state comprising a combination of double and quadruple quarks \cite{Fariborz:2022nws}. The $f_{0}(1710)$ meson, serving as the isovector partner of the $a_{0}(1817)$, does not rule out the possibility of being a scalar glueball. However, given the limited information available on the structure of the $a_{0}(1817)$, further resonance measurements are imperative. Consequently, the pressing task at hand involves detecting the $a_{0}(1817)$ in other experimental settings.

Upon consulting the Particle Data Group \cite{PDG}, we find that the $K^{-}p$ scattering experiment is particularly noteworthy. Since the discovery of $K$-mesons \cite{Rochester:1947mi}, kaon beams have naturally emerged as a powerful tool for exploring strange hadrons and hypernuclei \cite{Wang:2019uwk}. Experimental facilities such as J-PARC \cite{Jparc} and OKA@U-70 \cite{Obraztsov:2016lhp} offer excellent opportunities for such investigations.
Several literature sources have presented the production of the $a_0(980)$ in the reaction $K^- p \rightarrow \Lambda\eta \pi^{+}\pi^{-}$ \cite{Ammar:1968zur,Wells:1975tf,Amsterdam-CERN-Nijmegen-Oxford:1978mws,Flatte:1976xu}. The discovery of the $\phi(1020)$ meson has been achieved in the reaction $K^- p \rightarrow KKn$ \cite{Borenstein:1972sb}. Moreover, the $a_1(1260)$ and $D(1285)$ mesons have been observed in the reactions $K^{-} p \rightarrow \Sigma^{-} \pi^{+}\pi^{+}\pi^{-}$ \cite{Amsterdam-CERN-Nijmegen-Oxford:1977eoy} and $K^- p \rightarrow \Lambda\eta \pi^{+}\pi^{-}$ \cite{Amsterdam-CERN-Nijmegen-Oxford:1978mws}, respectively. These examples provide further support for the possibility of observing $a_{0}(1817)$ in $K^{-}p$ scattering experiments.
In our previous work, we successfully calculated the production of the $\phi(2170)$ meson via the reaction $K^- p \rightarrow \phi(2170)\Lambda$ \cite{Wang:2019uwk}, the $X_{0}(2900)$ state in $K^+ p \rightarrow \Sigma_{c}^{++}X_{0}(2900)$ \cite{Lin:2022eau}, and the $\eta_{1}(1855)$ meson through $K^{-}p\rightarrow \eta_{1}(1855)\Lambda$ \cite{Wang:2022sib} using efficient Lagrangian methods and the Regge trajectory model. The numerical results obtained from these calculations provide valuable insights for future experimental endeavors.
	
In this study, we explore the production mechanism of the scalar meson $a_{0}(1817)$ in $K^{-}p$ scattering utilizing an efficient Lagrangian approach, focusing on meson-induced reactions with $K$-meson exchange solely in the $t$-channel. Detailed information regarding our methodology will be presented in the subsequent section. The calculation of both the total cross section and differential cross section for the $K^{-}p\rightarrow
a_{0}(1817)\Lambda$ reaction holds significant relevance for future high-precision experimental investigations in this field.
	
This paper is structured as follows: In Section \ref{sec2}, we present the efficient Lagrangian method and the Regge trajectory model employed for the analysis of the $a_{0}(1817)$. The numerical results for the total and differential cross sections are presented in Section \ref{sec3}. Finally, we summarize our findings and draw conclusions in Section \ref{sec4}.

	\section{Formalism}
	\label{sec2}
	
\begin{figure}[htbp]

            \begin{center}
            	\includegraphics[scale=0.55]{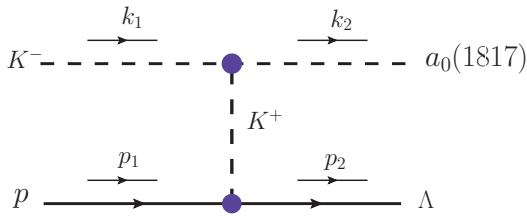}
            \end{center}
            \caption{Feynman diagram for the $K^{-}p\rightarrow \protect%
            	\ a _{0}(1817)\Lambda $ reaction.}
            \label{Fig: Feynman}
	\end{figure}

The production of the scalar meson $a_{0}(1817)$ through kaon-induced reactions on a proton target, with $t$ channel $K^{+}$ meson exchange, is illustrated in Fig.~\ref{Fig: Feynman}. In this study, we neglect the contribution from the $s$ channel with nucleon pole, as it is known to be negligibly small. Typically, the contribution of the $u$ channel with nucleon exchange is also minimal and can be neglected at low energies. Moreover, at high energies, the Reggeized treatment of the $u$ channel renders its contribution to the total cross section small and negligible. Therefore, we do not include the contributions from nucleon resonances in the $u$-channel in the current calculation.

 \begin{table} \small
	\caption{\label{tab:table1}
	The partial decay widths of the $\protect a _{0}(1817)$ predicted in Ref. \cite{Guo:2022xqu}. }
	\begin{ruledtabular}
		\begin{tabular}{ccccc}
			channel &
			 $\pi\eta $  &$\pi\eta^{\prime } $  & $\pi\eta(1295) $& $\pi\eta(1475)$ \\ \hline
			$\Gamma $ (MeV) &22.4$\sim$24.9&27.8$\sim$36.2&18.0$\sim$47.5&6.2$\sim$34.2\\ \hline
             channel  &$\pi b_{1}(1235) $ & $K\bar{K} $  & $\pi f_{1}(1285) $ &$\rho\omega  $ \\ \hline
             $\Gamma $ (MeV) &10.8$\sim$19.2&7.5$\sim$12.9&0.2$\sim$9.5&0.1$\sim$3.3\\
		\end{tabular}
	\end{ruledtabular}
\end{table}

	
Studying the strong interaction at the quark-gluon level in the energy region where the resonance can be detected poses significant challenges. Therefore, in this study, we employ the effective Lagrangian method to perform the necessary calculations. In the case of kaon-induced production of the $a_{0}(1817)$, the relevant Lagrangians for the $t$ channel are given by \cite%
	 {Wang:2015zcp,Ryu:2012tw,Liu:2008qx,Xiang:2020phx}%
	\begin{eqnarray}
		\mathcal{L}_{a_{0}KK} &=&\frac{f_{a_{0}KK}}{2m_{K}}a_{0}\partial _{\mu }%
		\overrightarrow{K}\cdot \partial ^{\mu }\overrightarrow{K}, \\
		\mathcal{L}_{KN\Lambda } &=&ig_{KN\Lambda }\bar{N}\gamma _{5}\Lambda K{\ +}%
		\text{ H.c.},
	\end{eqnarray}%
	where the $a_{0}$, ${K}$, $N$ and $\Lambda $ stand for the $a_{0}(1817)$, $K$%
	, nucleon and $\Lambda $ fields, respectively.
	
The coupling constant $g_{KN\Lambda }=-13.24$, which can be determined \cite{Wang:2015zcp} based on the SU(3) flavor symmetry relation \cite{Oh:2006hm,Oh:2006in}, plays a crucial role in our analysis. Additionally, the coupling constant $f_{a_{0}KK}$ can be determined from the decay width $\Gamma_ {a_{0}\rightarrow \bar{K}K}$, as indicated by the calculation results using the Nijmegen potential \cite{Stoks:1999bz}%
	\begin{eqnarray*}
		\Gamma _{a_{0}\rightarrow K^{+}K^{-}} &=&\frac{2}{3}\Gamma
		_{a_{0}\rightarrow \bar{K}K} \\
		&=&\left( \frac{f_{a_{0}KK}}{2m_{K}}\right) ^{2}\frac{%
			(M_{a_{0}}^{2}-2m_{K}^{2})^{2}}{32\pi M_{a_{0}}^{2}}|\vec{p}_{K}^{~\mathrm{%
				c.m.}}|  \label{AmpT1}
	\end{eqnarray*}%
	with%
	\begin{equation}
		|\vec{p}_{K}^{~\mathrm{c.m.}}|=\frac{\lambda
			^{1/2}(M_{a_{0}}^{2},m_{K}^{2},m_{K}^{2})}{2M_{a_{0}}}.
	\end{equation}%
	 Here, $\lambda$ represents the K\"{a}llen function, defined as $\lambda(x,y,z) = \sqrt{(x-y-z)^{2}-4yz}$. $M_{a_{0}}$ and $m_{K}$ denote the masses of $a_{0}(1817)$ and the kaon meson, respectively. By considering the decay width $\Gamma_{a_{0}\rightarrow K^{+}K^{-}}$ to be 5 MeV, we find that the corresponding coupling constant $f_{a_{0}KK}$ is determined to be 0.52.
	
	
	Based on the aforementioned Lagrangians, the amplitude for the production of the $a_{0}(1817)$ through $t$ channel $K^{+}$ exchange in $K^{-}p$ scattering can be expressed as follows:%
	\begin{eqnarray}
		\mathcal{M}_{K} &=&i\frac{f_{a_{0}KK}}{2m_{K}}g_{KN\Lambda }F(q^{2})\bar{u}%
		_{\Lambda }(p_{2})  \notag \\
		&&\times \gamma _{5}\frac{1}{t-m_{K}^{2}}(q_{\mu }\cdot k_{1}^{\mu
		})u_{N}(p_{1}).  \label{AmpT2}
	\end{eqnarray}%
	In the above expression, $\bar{u}_{\Lambda}$ and $u_{N}$ represent the Dirac spinors of the $\Lambda$ hyperon and nucleon, respectively. For the $t$ channel meson exchange \cite{Liu:2008qx}, a form factor $F(q^{2})=(\Lambda_{t}^{2}-m^{2})/(\Lambda_{t}^{2}-q^{2})$ is utilized. Here, $t=q^{2}=(k_{1}-k_{2})^{2}$ represents the Mandelstam variable. The parameter $\Lambda _{t}$, the only free parameter in the form factor, will be discussed in detail in Section \ref{sec3}.
	
	
	The Regge trajectory model has proven to be successful in analyzing hadron production at high energies \cite{Wang:2015xwa,Haberzettl:2015exa,Ozaki:2009wp,Wang:2017plf}. It provides a framework to study the spectral behavior of traditional light mesons \cite{Anisovich:2000kxa}. In this model, the Reggeization procedure is performed by replacing the $t$ channel propagator in the Feynman amplitudes (Eq. (\ref{AmpT2})) with the Regge propagator, which can be expressed as follows:
	\begin{equation}
		\frac{1}{t-m_{K}^{2}}\rightarrow \left( \frac{s}{s_{scale}}\right) ^{\alpha
			_{K}(t)}\frac{\pi \alpha _{K}^{\prime }}{\Gamma \lbrack 1+\alpha
			_{K}(t)]\sin [\pi \alpha _{K}(t)]}.
	\end{equation}%
	Here, the factor $s_{scale}$ is equal to 1 GeV. In addition, the Regge
	trajectory $\alpha _{K}(t)$ read as \cite{Wang:2015zcp},%
	\begin{equation}
		\alpha _{K}(t)=0.70(t-m_{K}^{2}).\quad \ \
	\end{equation}%
	It is note that no additional parameter is introduced after the Reggeized
	treatment applying.
	\section{Numerical results}
	\label{sec3}
	\subsection{Cross section}
	
	In the following calculations, we can determine the cross section of the $K^{-}p\rightarrow a_{0}(1817)\Lambda$ reaction. The differential cross section in the center of mass (c.m.) frame is given by:
	\begin{equation}
		\frac{d\sigma }{d\cos \theta }=\frac{1}{32\pi s}\frac{\left\vert \vec{k}%
			_{2}^{{~\mathrm{c.m.}}}\right\vert }{\left\vert \vec{k}_{1}^{{~\mathrm{c.m.}}%
			}\right\vert }\left( \frac{1}{2}\sum\limits_{\lambda }\left\vert \mathcal{M}%
		\right\vert ^{2}\right),
	\end{equation}%
	where the variable $s=(k_{1}+p_{1})^{2}$ represents the squared center of mass energy, and $\theta$ represents the angle between the outgoing $a_{0}(1817)$ meson and the direction of the kaon beam in the center of mass frame. $\vec{k}_{1}^{\mathrm{c.m.}}$ and $\vec{k}_{2}^{\mathrm{c.m.}}$ represent the three-momenta of the initial kaon beam and the final $a_{0}(1817)$ meson, respectively.

\begin{figure}[htbp]
	\centering
	\includegraphics[scale=0.45]{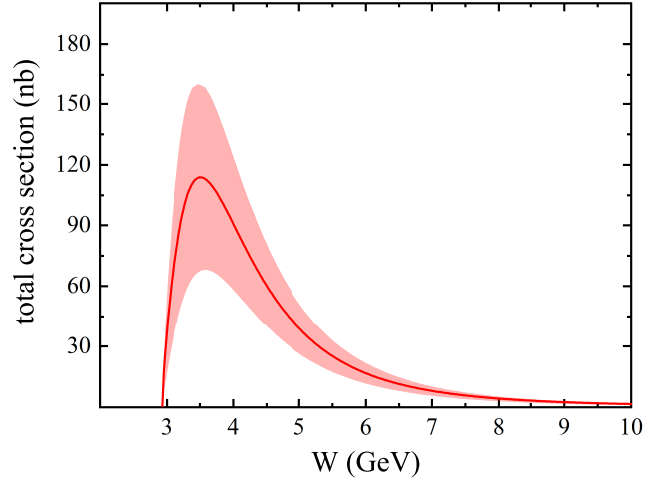}
	\caption{{The energy dependence of the total cross section for
			production of the $\protect a_{0}(1817)$ and through $t$ channel with cutoff $\Lambda _{t}=1.6\pm 0.3$ GeV. The
			Full (red)  line is for the  $K^{-}p\rightarrow \protect a _{0}(1817)\Lambda
			$ reaction. The bands stand for the error bar of cutoff $%
			\Lambda _{t}.$}}
	\label{fig:total}
\end{figure}

Since there is no available experimental data for the $K^{-}p\rightarrow a_{0}(1817)\Lambda$ reaction, we provide predictions for the cross section based on our calculations, as shown in Fig.~\ref{fig:total}. The cutoff parameter in the form factor serves as the only free parameter in these calculations. In previous studies, different values of the cutoff parameter have been used in related processes. For instance, in the $\pi^{-}p\rightarrow K^{\ast}\Sigma^{\ast}$ scattering process based on Reggeized $t$ channel $K^{(\ast)}$ exchange \cite{Xiang:2020phx}, a value of $\Lambda_{t}=1.67\pm 0.04$ GeV was employed. In another study \cite{Ozaki:2009wp}, to achieve better agreement with experimental data, a value of $\Lambda_{t}=1.55$ GeV was chosen for the Reggeized $t$ channel with $K$ and $K^{}$ exchange. Additionally, in the kaon-induced reaction $K^{-}p\rightarrow \eta_{1}(1855)\Lambda$, a cutoff value of $1.6\pm 0.3$ GeV for the $t$-channel $K^{+}$ exchange was considered \cite{Wang:2022sib}. In this work, we adopt $\Lambda_{t}=1.6\pm 0.3$ GeV to ensure a more reliable and feasible conclusion.

\begin{figure}
	\centering
	\includegraphics[scale=0.43]{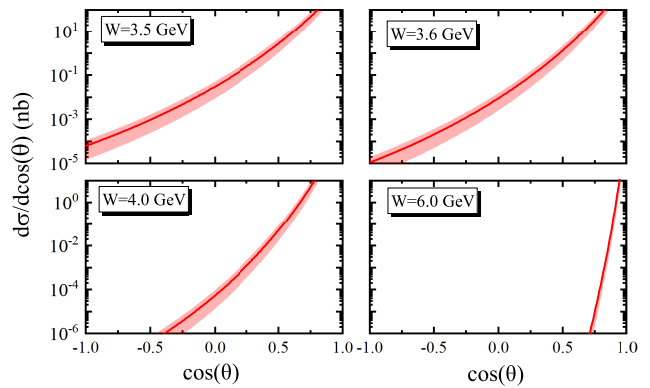}
	\caption{The differential cross section $d\protect\sigma %
		/d\cos \protect\theta $ of the $\protect a _{0}(1817)$ and $\protect a %
		_{0}(1817)$ production at different center-of-mass (c.m.) energies
		$W=3.5,3.6,4.0,6.0$  GeV.}
	\label{fig:dcs}
\end{figure}

In Fig. \ref{fig:total}, the total cross section exhibits a clear variation trend within the energy range of $W=2$ to $10$ GeV. Notably, there is a prominent peak between $W=3.4$ and $3.6$ GeV, indicating the potential for observing the $a_{0}(1817)$ resonance through $K^{-}p$ interactions within this energy range. The increase in the total cross section is steep leading up to the peak, with a value of $113$ nb at a center-of-mass energy of $3.5$ GeV. Following the peak, the downward trend becomes less pronounced. Taking into account the range of $\Lambda_{t}$ as $1.6\pm 0.3$ GeV and considering the error band, the total cross section varies by 67 nb from the value at $W=3.5$ GeV.

In Fig.~\ref{fig:dcs}, the predicted differential cross section of the $K^{-}p\rightarrow a_{0}(1817)\Lambda$ reaction is presented based on the Regge trajectory model, using a cutoff value of $\Lambda_{t}=1.6\pm 0.3$ GeV. It is evident from this work that the differential cross section is highly dependent on the scattering angle $\theta$. As the energy increases, the reaction exhibits a strong forward scattering and gradually strengthens. Therefore, the reinforcement treatment can be effectively validated through forward angle measurements.

\begin{figure}
	\centering
	\includegraphics[scale=0.42]{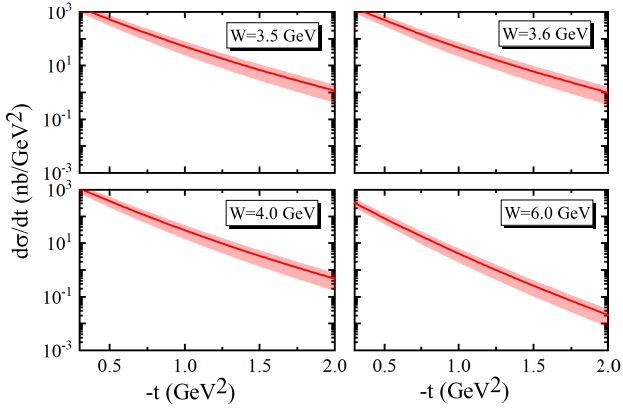}
	\caption{The $t$-distribution for the $K^{-}p\rightarrow
		\protect a _{0}(1817)\Lambda $ and $K^{-}p\rightarrow \protect a_{0}(1817)\Lambda $ reactions at different c.m. energies $%
		W=3.5$, $3.6$, $4.0$, $6.0$ GeV. Here, the notations are as that in Fig. \protect\ref%
		{fig:total}.}
	\label{fig:t-dcs}
\end{figure}

Figure \ref{fig:t-dcs} shows the $t$-distribution for the $K^{-}p\rightarrow a_{0}(1817)\Lambda$ reaction. It can be observed that the differential cross sections gradually decrease with increasing momentum transfer $t$. However, as $t$ becomes smaller, the differential cross section values continue to increase, and this phenomenon requires further experimental verification.

\subsection{Dalitz process}

\begin{figure}
	\centering
	\includegraphics[scale=0.42]{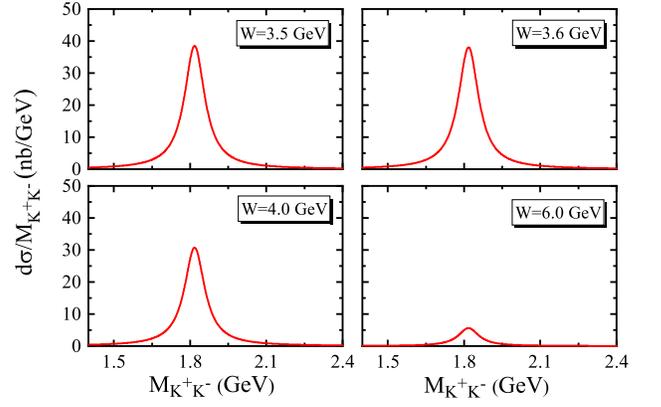}
	\caption{The invariant-mass distribution $d\protect\sigma %
	_{K^{-}p\rightarrow \protect a _{0} \Lambda \rightarrow \protect K^{+}K^{-} \Lambda }/ dM_{\protect K^{+}K^{-} }$
	reactions at different c.m. energies $W=3.5,3.6, 4.0, 6.0 $ GeV.}
	\label{fig:dalitz}
\end{figure}

\begin{figure}
	\centering
	\includegraphics[scale=0.42]{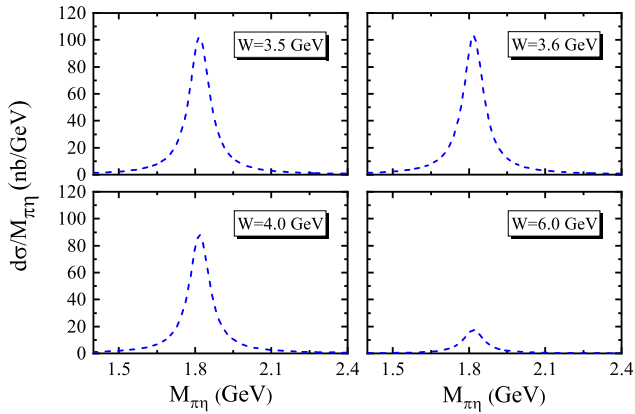}
	\caption{The invariant-mass distribution $d\protect\sigma %
		_{K^{-}p\rightarrow \protect a _{0}(1817) \Lambda \rightarrow \protect \pi\eta \Lambda }/ dM_{\protect \pi\eta }$
		reactions at different c.m. energies $W=3.5,3.6, 4.0, 6.0 $ GeV.}
	\label{fig:dalitz1}
\end{figure}

According to Table \ref{tab:table1}, it is evident that the $a_{0}(1817)$ meson frequently appears as an intermediate state in various decay processes. The minimum decay width for $a_{0}\rightarrow K\bar{K}$ is $\Gamma_{a_{0}\rightarrow K\bar{K}}=7.5$ MeV, while the minimum decay width for $a_{0}\rightarrow \pi\eta$ is $\Gamma_{a_{0}\rightarrow \pi\eta}=22.4$ MeV. In this study, we aim to calculate the Dalitz process for $K^{-}p\rightarrow a_{0}(1817)\Lambda \rightarrow K^{+}K^{-}\Lambda$ and $K^{-}p\rightarrow a_{0}(1817)\Lambda \rightarrow \pi\eta\Lambda$, respectively. The Dalitz process is of great importance and can provide valuable insights for future experimental investigations. Generally, the invariant mass distribution for the Dalitz process can be defined based on the two-body process \cite{Kim:2017nxg}
\begin{equation}
	\frac{d\sigma _{K^{-}p\rightarrow a _{0}\Lambda \rightarrow K^{+}K^{-} \Lambda }}{dM_{K^{+}K^{-} }}\approx \frac{2M_{a _{0}}M_{K^{+}K^{-}}}{\pi }\frac{\sigma _{K^{-}p\rightarrow a _{0}\Lambda
		}\Gamma _{a _{0}\rightarrow K^{+}K^{-} }}{(M_{K^{+}K^{-}
		}^{2}-M_{a _{0}}^{2})^{2}+M_{a _{0}}^{2}\Gamma
		_{a _{0}}^{2}},
\end{equation}%
\begin{equation}
	\frac{d\sigma _{K^{-}p\rightarrow a _{0}\Lambda \rightarrow \pi\eta \Lambda }}{dM_{\pi\eta  }}\approx \frac{2M_{a _{0}}M_{\pi\eta }}{\pi }\frac{\sigma _{K^{-}p\rightarrow a _{0}\Lambda
		}\Gamma _{a _{0}\rightarrow \pi\eta  }}{(M_{\pi\eta
		}^{2}-M_{a _{0}}^{2})^{2}+M_{a _{0}}^{2}\Gamma
		_{a _{0}}^{2}}.
\end{equation}%
          Here, the total width of $a_{0}$ meson, denoted as $\Gamma_{a_{0}}$, is 97 MeV. For the partial width $\Gamma_{a_{0}\rightarrow K^{+}K^{-}}$, we consider a value of 5 MeV, and for $\Gamma_{a_{0}\rightarrow \pi\eta}$, we use the value of 22.4 MeV. Based on these parameters, we calculate the invariant-mass distributions $d\sigma_{K^{-}p\rightarrow a_{0}\Lambda \rightarrow K^{+}K^{-}\Lambda}/dM_{K^{+}K^{-}}$ and $d\sigma_{K^{-}p\rightarrow a_{0}\Lambda \rightarrow \pi\eta\Lambda}/dM_{\pi\eta}$ for center-of-mass energies ranging from $W=3.5$ GeV to $W=6$ GeV. The results are shown in Fig. \ref{fig:dalitz} and Fig. \ref{fig:dalitz1}. It can be observed from these figures that there is a peak near the center-of-mass energy of approximately 1.82 GeV, which has direct implications for the experimental detection of the $a_{0}(1817)$.

To further assess the feasibility of detecting $a_{0}(1817)$ in $K^{-}p$ interactions, we calculate the ratio $\sigma(K^{-}p\rightarrow a_{0}(1817)\Lambda \rightarrow K^{+}K^{-}\Lambda) / \sigma(K^{-}p\rightarrow K^{+}K^{-}\Lambda)$. In Fig.~\ref{fig:total}, the cross section for $a_{0}(1817)$ production in $K^{-}p$ scattering is estimated to be approximately 110 nb at $W=3.37$ GeV. Assuming a branching ratio of $BR(a_{0}(1817) \rightarrow K^{+}K^{-}) \approx 5.2\%$, we obtain a total cross section of $\sigma_{K^{-}p\rightarrow a_{0}(1817)\Lambda \rightarrow K^{+}K^{-}\Lambda} \approx 5.72$ nb at $W=3.37$ GeV.

In Ref. \cite{Mott:1969hwi}, a total cross section of 35 $\mu$b is reported for $W=3.37$ GeV. Based on this value, the ratio at $W=3.37$ GeV can be calculated as follows:
     \begin{equation}
     	\frac{\sigma(K^{-}p\rightarrow a _{0}(1817)\Lambda \rightarrow K^{+}K^{-}\Lambda )}{\sigma(K^{-}p\rightarrow K^{+}K^{-}\Lambda)} \approx 0.016\%.
     \end{equation}%

Considering the current experimental landscape, we are optimistic about the potential of the J-PARC experiment in detecting the $a_{0}(1817)$ in $K^{-}p$ scattering \cite{Nagae:2008zz, Jparc}. The experimental conditions at J-PARC are well-suited for this purpose. Based on the specifications of the J-PARC experiment, it is estimated that approximately 42,000 events of $K^{+}K^{-}\Lambda$ can be generated in 100 days, among which about several events are expected to involve the $a_{0}(1817)$. By performing calculations, we find that at $W=3.37$ GeV, the cross section for $K^{-}p\rightarrow a_{0}(1817)\Lambda \rightarrow \pi\eta\Lambda$ is approximately 25.40 nb, considering a branching ratio of $BR(a_{0}(1817) \rightarrow \pi\eta) \approx 23.1\%$. These events can be reliably detected at J-PARC every 100 days, with dozens of events specifically related to the $a_{0}(1817)$. Consequently, the $a_{0}(1817)$ can be confidently observed from the $K^{-}p\rightarrow a _{0}(1817)\Lambda \rightarrow \pi\eta\Lambda$ reaction under the current experimental conditions.
Therefore, with the future upgrade of J-PARC, there is a promising opportunity to discover and study the $a_{0}(1817)$ in greater detail.


\section{SUMMARY}
\label{sec4}
In the past two years, significant progress has been made in the study of the isovector scalar meson $a_{0}(1817)$ by the BaBar and BESIII Collaborations. However, the measured resonance parameters of the $a_{0}(1817)$ differ between these experiments, and there is still a lack of sufficient data to fully understand its structure \cite{PDG}. In order to further investigate the intrinsic properties of the $a_{0}(1817)$, we propose to explore its characteristics through $K^{-}p$ scattering.

There are several reasons for choosing the $K^{-}p$ interaction to search for the $a_{0}(1817)$. Firstly, the ground state particle of the $a_{0}(1817)$, $a_{0}(980)$, has already been observed in the $K^{-}p$ scattering process. Secondly, the decay channel $a_{0}(1817) \rightarrow K^{+}K^{-}$ plays a significant role in the overall decay of the $a_{0}(1817)$. By employing the effective Lagrangian method and the Regge trajectory model in quantum field theory, we calculate the total and differential cross sections of $K^{-}p\rightarrow a_{0}(1817) \Lambda$.

Our results indicate that the total cross section exhibits a peak at a center-of-mass energy of $W=3.4 \sim 3.6$ GeV, suggesting that this energy range is ideal for detecting the $a_{0}(1817)$ through the $K^{-}p\rightarrow a_{0}(1817)\Lambda$ reaction. Moreover, the differential cross section is highly sensitive to the scattering angle $\theta$ and the minimum momentum transfer $t$. Therefore, high-precision data from experimental facilities such as J-PARC, OKA@U-70, and SPS@CERN, which provide suitable kaon beams, are eagerly awaited.

Based on the current experimental conditions, the detection of the $a_{0}(1817)$ from $K^{-}p\rightarrow \pi\eta\Lambda$ is considered more feasible compared to $K^{-}p\rightarrow K^{+}K^{-}\Lambda$. The theoretical insights obtained in this study will provide valuable information for future experiments aimed at identifying and characterizing the $a_{0}(1817)$ state.

\section*{ACKNOWLEDGMENTS}
This work is supported by the National Natural Science Foundation of China under Grants No. 12065014, No. 12047501 and No. 12247101, and by the Natural Science Foundation of Gansu province under Grant No. 22JR5RA266. We acknowledge the West Light Foundation of The Chinese Academy of Sciences, Grant No. 21JR7RA201.
X. L. is also supported by the China National Funds for Distinguished Young Scientists under Grant No. 11825503, National Key Research and Development Program of China under Contract No. 2020YFA0406400, the 111 Project under Grant No. B20063, the Fundamental Research Funds for the Central Universities, and the project for top-notch innovative talents of Gansu province.

\end{document}